%% file: main.tex
\documentclass[5p,twocolumn]{elsarticle}

\usepackage{amssymb,epsfig,graphicx,amssymb,subeqnarray, multirow, pstricks,colortbl, bm}
\usepackage{units,tabularx,array,booktabs,booktabs,algorithm, algorithmic}
\usepackage[cmex10]{amsmath}
\usepackage[tight,footnotesize]{subfigure}  
\usepackage{soul, verbatim}
\include{macpap2}
\PassOptionsToPackage{hyphens}{url}\usepackage{hyperref}
\usepackage[]{xcolor}
\usepackage{paralist}

\journal{Alzheimer's and Dementia}

\begin{document}

\begin{frontmatter}

\title{Random forest prediction of Alzheimer's disease using \\pairwise selection from time series data.}

\author{P.J. Moore\corref{cor1}\fnref{fn1}}
\ead{moorep@maths.ox.ac.uk}
\cortext[cor1]{Corresponding author}
\fntext[fn1]{Postdoc, Mathematical Institute, University of Oxford. } 

\author{J. Gallacher\fnref{fn2}}
\fntext[fn2]{Professor, Department of Psychiatry, University of Oxford.}

\author{T.J. Lyons \fnref{fn3}}
\fntext[fn3]{Professor, Mathematical Institute, University of Oxford.}

\author{\\for the Alzheimer’s Disease Neuroimaging Initiative \fnref{fn4}}
\fntext[fn4]{Data used in preparation of this article were obtained from the Alzheimer’s Disease
Neuroimaging Initiative (ADNI) database (adni.loni.usc.edu). As such, the investigators
within the ADNI contributed to the design and implementation of ADNI and/or provided data
but did not participate in analysis or writing of this report. A complete listing of ADNI
investigators can be found at:\url{http://adni.loni.usc.edu/}.}

\address{University of Oxford}


\begin{abstract}
Time-dependent data collected in studies of Alzheimer's disease usually has missing and irregularly sampled data points.  For this reason time series methods which assume regular sampling cannot be applied directly to the data without a pre-processing step.  In this paper we use a machine learning method to learn the relationship between pairs of data points at different time separations.  The input vector comprises a summary of the time series history and includes both demographic and non-time varying variables such as genetic data. The dataset used is from the 2017 TADPOLE grand challenge which aims to predict the onset of Alzheimer's disease using including demographic, physical and cognitive data. The challenge is a three-fold diagnosis classification into AD, MCI and control groups, the prediction of ADAS-13 score and the normalised ventricle volume.  While the competition proceeds, forecasting methods may be compared using a leaderboard dataset selected from the Alzheimer's Disease Neuroimaging Initiative (ADNI) and with standard metrics for measuring accuracy. For diagnosis, we find an mAUC of 0.82, and a classification accuracy of 0.73.  The results show that the method is effective and comparable with other methods.
\end{abstract}


\end{frontmatter}


%
\section{Introduction}
Alzheimer's disease (AD) is an irreversible brain disorder which progressively affects cognition and behaviour, and results in an impairment in the ability to perform daily activities.  It is the most common form of dementia in older people, affecting about 6\% of the population aged over 65, and it increases in incidence with age. The initial stage of AD is characterised by memory loss, and this is the usual presenting symptom.  Memory loss is one constituent of mild cognitive impairment (MCI), which can be an early sign of Alzheimer's disease.  MCI is diagnosed by complaints of subjective memory loss (preferably corroborated by a close associate or partner of the individual), impairment of memory function, unimpaired general cognition and behaviour but with no evidence of dementia \cite{nestor2004}.  MCI does not always progress to dementia or to a diagnosis of Alzheimer's disease, but those with amnestic mild MCI, the type of MCI characterised by memory impairment, are more likely to develop dementia than those without this diagnosis.  In cases where an individual does develop Alzheimer's disease, the phase of MCI ends with a marked decline in cognitive function, lasting two to five years, in which semantic memory (the recall of facts and general knowledge) and implicit memory (the long-term, nonconscious memory evidenced by priming effects) also becomes degraded.  
\nind
Clinical diagnosis of dementia relies on information from a close associate or partner of the individual, and on cognitive and physical examinations.  Once dementia is diagnosed it is usually subclassified into Alzheimer's disease, vascular dementia or Lewy Body dementia \cite{burns2009}\cite{dubois2007}, these three classes making up the majority of cases.  Risk factors for Alzheimer's disease are multifarious, including sociodemographic (in particular age), genetic (notably ApoE status), and medical history (such as a diagnosis of depression).  The cause of Alzheimer's disease is not fully understood, but plaques containing amyloid $\beta$--peptide (A$\beta$) in brain tissue and neurofibrillary tangles containing tau protein are the primary histological features\cite{tiraboschi2004}. 

%
\subsection{Predicting Alzheimer's disease}
The disease pathology leads to an progressive, irreversible loss of brain function which suggests that prospective drug therapies should be tested for efficacy as early in the process as possible.  So there has been a demand for predicting which individuals will develop AD as early as possible in order to test drug therapies which might inhibit or prevent tissue damage.  There has been much research effort put into the prediction of an AD diagnosis among those who are diagnosed with MCI, in particular using imaging to detect early signs of the disease pathology: a meta analysis of 32 structural MRI or amyloid PET imaging studies that reported conversion to AD in patients with MCI is given by Seo \etal \cite{seo2017}. 
This analysis concluded that amyloid PET is a better predictor of progression to AD from MCI than MRI atrophy measures (effect size 1.32 vs 0.77), but that MRI on entorhinal cortex atrophy (effect size 1.26) is comparable in prediction value to that of amyloid PET.  Another comparison of biomarker predictivity found that the highest predictive accuracy was achieved by combinations of amyloidosis and neurodegeneration biomarkers \cite{prestia2015}.  The individual biomarker with the best performance was [18F]-fluorodeoxyglucose-positron emission tomography (FDG-PET) which measures temporoparietal hypometabolism.  
\nind
Cognitive markers have also been widely applied for early detection of Alzheimer's disease: a review is presented by Gainotti \etal in \cite{gainotti2014} which concluded that measures of delayed recall are the best neuropsychological markers of conversion from MCI to AD.  Significantly, they also suggest that MCI subjects with deficits in multiple cognitive domains including memory may not be the best candidates for clinical trials of disease-modifying drugs: of this group about 50\% of this will convert to AD within 2 years,  making their condition less modifiable than for those who are at an earlier stage of the disease.
\subsection{The TADPOLE challenge}
In evaluating a method for predicting Alzheimer's disease it is important to compare the results with the current state of the art in order to calibrate the accuracy.  In the past few years there have been a number of challenges which allow comparison between methods using a common data set and standardised evaluation metrics.  The CADDementia challenge \cite{bron2015standardized} compares algorithms for multi-class classification of AD, MCI and controls based on structural MRI data.  The Kaggle Neuroimaging challenge \url{https://www.kaggle.com/c/mci-prediction} \cite{sarica2018editorial} is based on the Kaggle machine learning platform and uses data from the Alzheimer’s Disease Neuroimaging Initiative (ADNI) \url{http://adni.loni.usc.edu/} which is one of the most commonly used data sets for studies of Alzheimer's disease\cite{weiner2017recent}.  The challenge involved a four-fold classification into AD, MCI, MCI converters to AD, and controls.  
\nind
The 2017 TADPOLE grand challenge \url{https://tadpole.grand-challenge.org/} is currently taking place with the evaluation to be completed by January 2019.  The challenge is a three-fold diagnosis classification into AD, MCI and control groups, and the prediction of ADAS-13 score and normalised brain volume \cite{tadpole}.  The TADPOLE challenge has the aim of predicting the onset of Alzheimer's disease using different modes of measurement, including demographic, physical and cognitive data. In common with many other studies\cite{weiner2017recent}, the TADPOLE data set is also derived from ADNI.  ADNI itself is comprised of four phases: ADNI-1 (2004), ADNI-GO (2009), ADNI-2 (2011), and ADNI-3 (2016).  ADNI-1 registered 200 healthy elderly, 400 participants with MCI, and 200 participants with AD, and the subsequent phases continued to add participants.  The TADPOLE competition involves predicting future data collected as part of the ADNI-3 phase. The competition organisers provide a leaderboard dataset which is separate from the main competition and which allows prediction methods from different teams to be evaluated.  The results presented here are derived from the TADPOLE leaderboard dataset. Since TADPOLE data is based on ADNI, data used in the preparation of this article were obtained from the (ADNI) database (\url{adni.loni.usc.edu}). ADNI is led by Principal Investigator Michael W. Weiner, MD. For up-to-date information see \url{www.adni-info.org}.  

%
\section{Forecasting}

%
\subsection{Leaderboard data}

\begin{figure}[htp]
\centering
\includegraphics[trim = 0mm 0mm 0mm 0mm, clip, width=8cm]{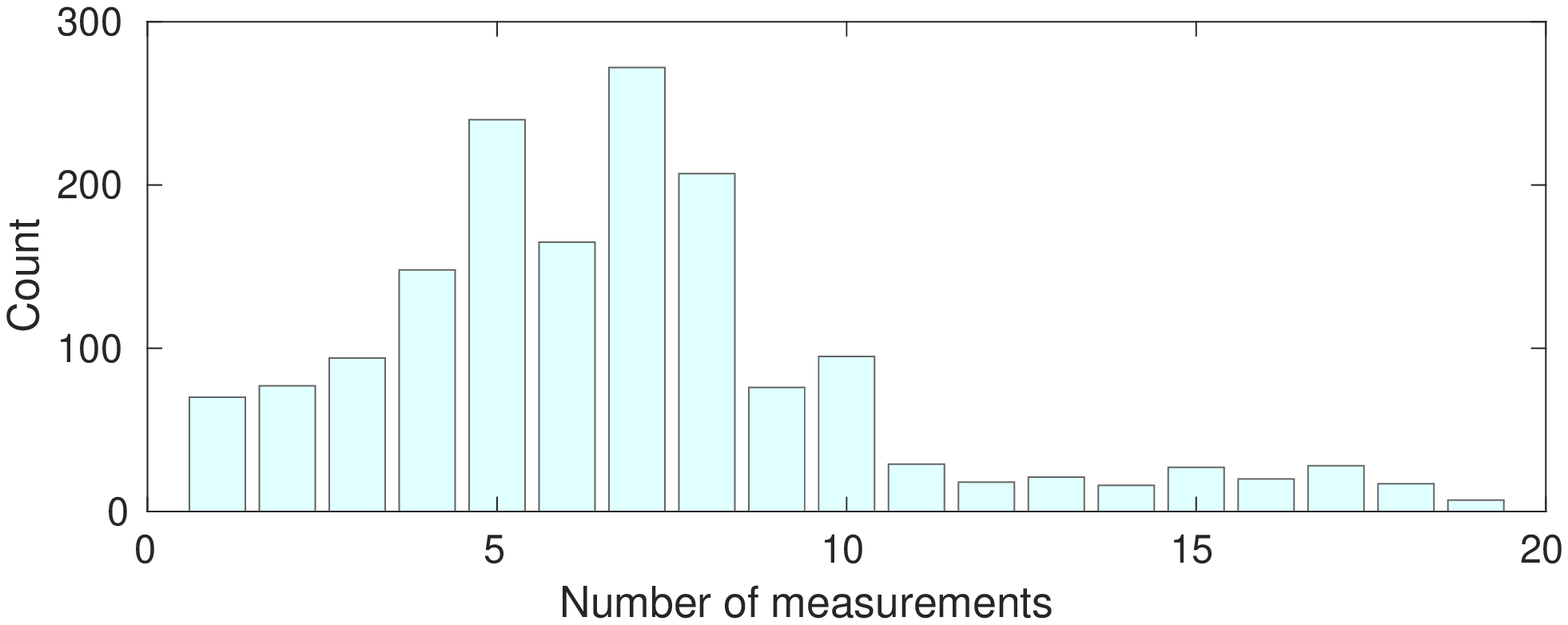} 
\includegraphics[trim = 0mm 0mm 0mm 0mm, clip, width=8cm]{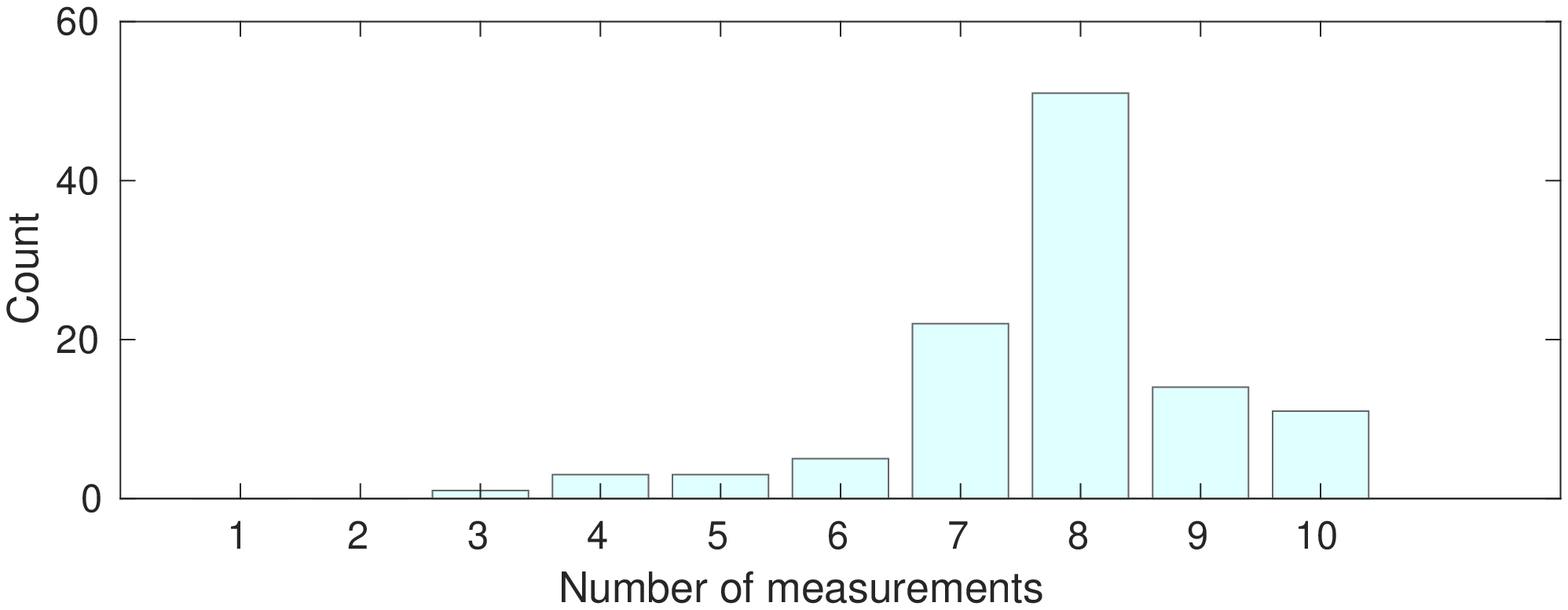} 
\caption[]{Histograms of time series lengths. Upper: training set LB1 whose time series may cover ADNI, ADNI-GO and ADNI-2.  Lower: Set LB2 which is formed only from time series in from ADNI-1.} 
\label{fig:bar_training_lengths}
\end{figure}

The leaderboard dataset has a training set LB1 and a set LB2 whose participants subsequently continued from ADNI-1 into ADNI GO/ADNI-2 to form the leaderboard test set LB4.  LB2 is formed of ADNI-1 time points for 110 participants who who were not diagnosed with AD at the last ADNI-1 time point.  The test set LB4 comprises the data points for those same LB2 participants during their continuing participation in ADNI GO/ADNI-2.  The training set LB1 comprises data from participants who are not represented in LB2.  The task is to predict the diagnosis, ADAS-13 score and the normalised ventricle volume for the set LB4 using set LB1 and the participant histories recorded in LB2.  The results are evaluated by comparison with LB4 data using a variety of metrics.   No information from LB4 may be used for model training, but demographic and other details about the participants who contributed to LB4 are available from LB2, and past time varying data such as imaging and cognitive measurements are also available from LB2.  A histogram of time series lengths for LB1 and LB2 is shown in Figure \ref{fig:bar_training_lengths}.

%

Features for prediction are selected from the demographic, cognitive and physical data variables in the ADNI/TADPOLE data.  The physical data comprises, among other measurements, MRI data (volumes, cortical thickness, surface area), PET (FDG, AV45 and AV1451), DTI (regional means of standard indices) and levels of markers from cerebral spinal fluid (CSF).

%

\subsection{Evaluation set}
For the purposes of training we form an evaluation set which is similar to the time series used in the leaderboard evaluation, that is LB2 from the ADNI-1 phase and the test set LB4 from the ADNI-GO and ADNI-2 phases.  We select the evaluation set from LB1 by choosing participants who match those of LB2 and whose ADNI-1 time series length is similar.  The post-ADNI-1 phases of this matched evaluation set can be used to assess the prediction accuracy, which should be similar to that of the test set LB4.  To create the evaluation set we examine each participant time series in LB2 and find those participants in LB1 who have a matching gender, ApoE status and age (to within 5 years) and whose diagnosis matches at the start and end of the ADNI-1 period.  If more than one matching participant is found we select the one who has the closest match for the time series length in the ADNI-1 phase.   The demographic characteristics and ApoE status of the participants from set LB2 and the matched evaluation set are shown in Figure \ref{tab:matched_set}.

%
\begin{table}[htp]
\small
\newcolumntype{x}[1]{>{\raggedleft\arraybackslash\hspace{0pt}}p{#1}}
\newcolumntype{P}{p{25mm}}
\newcolumntype{Q}{p{15mm}}

\centering
\begin{tabularx}{0.5\textwidth}{x{20mm}PP}
 & \it{LB2} &  \it{Matched set} \\
\midrule
\it{n:} & 110 & 92 \\         
\\
\it{Age:} & 59.9 (75.1) 87.9 & 57.8 (75.3) 84.8  \\
\\
\it{Male:} & 60.9\% & 60.9\% \\
\it{Female:} & 39.1\% & 39.1\% \\
\\
\it{APOE 0:} & 70.0\% & 68.5\% \\
\it{APOE 1:} & 27.3\% & 29.4\% \\
\it{APOE 2:} & 2.7\% & 2.2\% \\
\midrule

\end{tabularx}
\caption[]{Demographic characteristics and ApoE status for participants from the set LB2 compared with a matched evaluation set selected from LB1. The variables shown are the sample size $n$, age as minimum, mean and maximum, gender and ApoE status.}
\label{tab:matched_set}
\end{table}

%

\subsection{Training}
We train the model using the time series from LB1 and LB2.  Points from ADNI-GO and ADNI-2 which are used to compute prediction accuracy are not used for for training.  Training is performed on 90\% of the participants in the evaluation set with  accuracy measured on the remaining 10\%, this repeated for 10 independent splits of the whole evaluation set. We train only on time series with at least 4 points, this minimum having been determined during training.  The prediction variables are selected manually by optimising the mean accuracy found on the evaluation set.


%
\subsection{Forecasting method}

The purpose of a forecasting method is to predict a patient's condition at points in the future using demographic, cognitive and physical data variables from time points in the participant's history.  A common approach to automatic prediction is to use time series methods which use weighted combinations of past data points to predict the next data point.  Time series models in general encode a mapping from an $r$-dimensional space to the output, where time is not one of the input dimensions.  But many time series in the training data are short and the sampling periods are irregular, so much of the information in the training set lies in the mapping from the time delay between measurements to the output rather than in the sequence of input values\footnote{Irregular sampling and missing data can be managed by interpolation or by using appropriate methods such as Gaussian process regression \cite{rasmussen2006gaussian}, but these approaches entail making assumptions about the distributions.}.    Another approach is to use an input space formed of demographic variables $\lambda$, last diagnosis $g_{(t-\Delta t)}$, time since last measurement $\Delta t$, and map vectors in this to the output variable $g_t$.  Again assuming additive error, the model is,
\be g_t = f(\lambda,\, g_{(t-\Delta t)} \,,\Delta_t) +  \eps_t \ee
\nind
We use an ensemble of decision trees to estimate the regression function $f$.  Decision trees approximate the regression function by partitioning the input (feature) space into a set of rectangles \cite[p305]{hastie}.  The training algorithm iterates over all the features and selects the feature and split point that gives the best partition for the training data; this is repeated until a stopping criterion is met, such as a minimum number of points in the rectangle.  The best partition is that that which gives the minimum total impurity in the two subsets that are formed.  However, in our experience decision trees own tend to overfit the training data and perform poorly on new data, so we use a ensemble of trees and poll the individual results as an estimator.  Further improvement is seen with the random forest algorithm in which the features are chosen randomly at each split point \cite{breiman2001random}.  An introduction to the theory of random forests is given in Hastie \etalns\cite[ch16]{hastie}.

%
\section{Results}
There are three target outcomes for prediction: 1) Diagnosis, 2) the ADAS-13 score, and 3) VENTS-ICV which is the ventricles volume divided by intracranial volume.  We first present results from the evaluation set, followed by the test set accuracy. 
%

\subsection{Evaluation set}
%

\paragraph{Diagnosis}
The accuracy for ten independent splits of 10\% of the evaluation set, and the training error on the full evaluation set is shown in Table \ref{tab:results_diag}.

%
\begin{table}[htp]
\footnotesize
\newcolumntype{x}[1]{>{\raggedleft\arraybackslash\hspace{0pt}}p{#1}}
\newcolumntype{P}{p{7mm}}
\definecolor{LightGrey}{rgb}{0.4,0.4,0.4}
\centering
\begin{tabularx}{0.95\textwidth}{x{10mm}cccc}
   & \it{Acc(baseline)} & \it{Acc} & \it{mAUC} & \it{BCA} \\
\cmidrule(lr){2-5}
\multirow{10}{*}{\emph{Split:}}
& 61.54 & 69.23 & 0.87 & 0.75 \\
& 73.53 & 73.53 & 0.77 & 0.64 \\
& 69.57 & 69.57 & 0.73 & 0.44 \\
& 52.38 & 52.38 & 0.74 & 0.59 \\
& 86.49 & 86.49 & 0.83 & 0.78 \\
& 73.68 & 86.84 & 0.99 & 0.93 \\
& 70.21 & 74.47 & 0.85 & 0.78 \\
& 62.50 & 78.13 & 0.73 & 0.69 \\
& 91.43 & 91.43 & 0.76 & 0.78 \\
& 88.64 & 88.64 & 0.90 & 0.78 \\

\\
\it{Mean:} & 73.00 & 77.07 & 0.82 & 0.71 \\
\cmidrule(lr){2-5}
 
\it{Full:} &    73.25 &  76.50 &   0.85  &  0.77\\
\cmidrule(lr){2-5}

\end{tabularx}
\caption[]{Prediction accuracy for diagnosis in the evaluation set.  The first ten rows show the results for each partition of the evaluation set. The final row labelled \emph{Full} shows the training accuracy for the entire set of 92 participants.  The first column shows a baseline accuracy which is obtained by using the last diagnosis as the predictor for subsequent values.  The metrics used are taken from the TADPOLE competition and the codes reimplemented in order to double-check the results. As a measure of accuracy we report the multiclass area under the receiver operating curve (mAUC) and balanced classification accuracy (BCA).  The mAUC is based on Hand and Till's extension of AUC to multiple classes \cite{hand2001simple} which takes the average AUC over all the pairs of classes, where each pair of AUCs $\hat{A}(c_i,c_j), \hat{A}(c_j,c_i)$ is itself averaged.  The balanced classification accuracy (BCA) is based on point scores, and is the mean of the True Positive rate and True Negative rate.   }
\label{tab:results_diag}
\end{table}

%

\paragraph{ADAS-13 and VENTS-ICV}
The prediction error for ten independent splits of 10\% of the evaluation set, and the training error on the full evaluation set is shown in Table \ref{tab:results_adas13} where the first two columns show ADAS-13 results and the next two columns show those for the ratio of ventricle volume divided by intracranial volume.  
%

\begin{table}[htp]
\footnotesize
\newcolumntype{x}[1]{>{\raggedleft\arraybackslash\hspace{0pt}}p{#1}}
\centering
\begin{tabularx}{0.95\textwidth}{x{10mm}cccc}
 & \multicolumn{2}{c}{\emph{ADAS-13}} & \multicolumn{2}{c}{\emph{VENTS-ICV}}\\ 
  & \it{MAE(baseline)} & \it{MAE}  & \it{MAE(baseline)} & \it{MAE}  \\
\cmidrule(lr){2-5}
\multirow{10}{*}{\emph{Split:}}
& 9.16 & 6.18 & 0.0034 & 0.0011 \\
& 6.28 & 5.18 & 0.0027 & 0.0036 \\ 
& 5.11 & 5.15 & 0.0033 & 0.0017 \\ 
& 8.15 & 6.39 & 0.0037 & 0.0020 \\ 
& 5.06 & 4.43 & 0.0028 & 0.0017 \\ 
& 8.10 & 5.80 & 0.0041 & 0.0013 \\ 
& 8.76 & 6.05 & 0.0048 & 0.0026 \\ 
& 7.55 & 5.29 & 0.0027 & 0.0040 \\ 
& 4.27 & 5.61 & 0.0042 & 0.0022 \\ 
& 7.50 & 5.50 & 0.0023 & 0.0011 \\

\\
\it{Mean:} & 6.99 & 5.56 & 0.0034 & 0.0021 \\ 
\cmidrule(lr){2-5}
 
\it{Full:} & 6.97 & 5.71 & 0.0035 & 0.0021 \\ 
\cmidrule(lr){2-5}

\end{tabularx}
\caption[]{Mean absolute error for predicting ADAS-13 and VENTS-ICV in the evaluation set.  In each case the baseline is a prediction using the last value of the variable to be predicted.  The first ten rows show the results for each partition of the evaluation set. The final row labelled \emph{Full} shows the result for the entire set of 92 participants.   }
\label{tab:results_adas13}
\end{table}

%

\paragraph{Predictors}
The predictor variables that were selected during training are shown in Table \ref{tab:features}.  These were chosen by starting from a base set of variables and adding variables to increase the prediction accuracy.  The set of predictors for diagnosis and ADAS-13 are the same except that for ADAS-13 the most recent value of the variable \texttt{ADAS13} and its slope are added.  

%

\begin{table}[htp]
\small
\newcolumntype{Y}{>{\centering\arraybackslash}X}
\centering
\begin{tabularx}{0.5\textwidth}{p{23mm}l}
Variable & Meaning\\
\cmidrule(lr){1-2}

\texttt{RID} & Participant identifier\\
\texttt{TIME\_DELAY} & Number of months delay \\
\texttt{DX} & Diagnosis (NL or MCI or AD) \\
\texttt{AGE} & Age \\
\texttt{PTGENDER} & Gender \\
\texttt{ApoE} & ApoE allele\\
\texttt{FUSIFORM\_BL} & Fusiform volume at baseline \\
\texttt{MMSE} & MMSE Mini-mental state examination\\
\texttt{CDRSB} & CDRSB \\
\texttt{FAQ} & Functional activities questionnaire \\
\cmidrule(lr){1-2}
$\Delta$\texttt{VENTRICLES} & Ventricles volume slope\\
$\Delta$\texttt{HIPPOCAMPUS} & Hippocampus volume slope \\
$\Delta$\texttt{MMSE} & MMSE slope\\
\cmidrule(lr){1-2}
\end{tabularx}
\caption[]{The set of variables from which features are chosen for predicting diagnosis and ADAS-13 values.  The most recent value of each variable is used for prediction.  The \texttt{MMSE} value is thresholded at 26.  For predicting ADAS-13, the variables \texttt{ADAS13} and $\Delta$\texttt{ADAS13} are added.  For predicting \texttt{VENTS-ICV}, the two target variables \texttt{VENTS} and \texttt{ICV} are predicted separately using the last value and slope of the respective variable, with \texttt{TIME\_DELAY} also used.    The prefix $\Delta$ indicates that the slope of the variable over the history is used for prediction.}
\label{tab:features}
\end{table}
  

%

\subsection{Test set LB4}
The accuracy on the test set LB4 is shown as the highlighted row of Table \ref{tab:leaderboard} which reproduces the results from the TADPOLE competition leaderboard table.  The leaderboard shows entries in rank order where the rank is determined by the lowest sum of individual ranks for mAUC, ADAS-13 MAE and VENTS-ICV MAE.  For diagnosis the accuracy is mAUC=0.82 and BCA=0.73, which compares a mean accuracy on our evaluation set of mAUC=0.82 and BCA=0.71.  The similarity of test and evaluation errors shows that the evaluation set is well matched to the test set and that the model has not been overtrained.  For ADAS-13 prediction the test set error is MAE=5.19, and the evaluation set error is MAE=5.56 and for VENTS-ICV the test error is MAE=0.0023 and the evaluation error is MAE=0.0021.  The confusion matrix for the test set is shown in Table \ref{tab:confusion_matrix}.

%

\begin{table}[htp]
\footnotesize
\newcolumntype{Y}{>{\centering\arraybackslash}X}
\newcolumntype{P}{p{10mm}}
\definecolor{LightCyan}{rgb}{0.88,1,1}
\centering
\begin{tabularx}{0.75\textwidth}{p{25mm}Pccc}
& & \multicolumn{3}{c}{\emph{Predicted diagnosis}}\\ 

&  & NL & MCI & AD \\
\cmidrule(lr){2-5}
\multirow{3}{*}{\emph{Actual diagnosis:}}
 & NL & 193 & 2 & 0 \\
 & MCI & 54 & 88 & 8 \\
 & AD & 6 & 45 & 21 \\
\cmidrule(lr){2-5}

\end{tabularx}
\caption[]{Confusion matrix for predicting the diagnosis of the test set. In this case, the a point forecast is derived from the probabilistic forecast by selecting the diagnosis with the highest score.} 
\label{tab:confusion_matrix}
\end{table}


%

\begin{table}[htp]
\footnotesize
\newcolumntype{Y}{>{\centering\arraybackslash}X}
\newcolumntype{P}{p{7mm}}
\definecolor{LightCyan}{rgb}{0.88,1,1}
\centering
\begin{tabularx}{0.8\textwidth}{lPPPPPPP}
 \multicolumn{2}{c}{Diagnosis} & \multicolumn{3}{c}{ADAS-13} & \multicolumn{3}{c}{VENTS-ICV}  \\ 
\cmidrule(lr){1-2}\cmidrule(lr){3-5}\cmidrule(lr){6-8}
mAUC & BCA & MAE & WES & CPA & MAE & WES & CPA \\
\cmidrule(lr){1-8}

0.91 & 0.83 & 3.62 & 3.62 & 0.11 & 0.0020 & 0.0018 & 0.13 \\
0.93 & 0.85 & 3.72 & 3.10 & 0.02 & 0.0020 & 0.0016 & 0.15 \\
0.93 & 0.85 & 3.72 & 3.10 & 0.02 & 0.0020 & 0.0016 & 0.15 \\
0.91 & 0.83 & 3.67 & 3.67 & 0.12 & 0.0024 & 0.0022 & 0.08 \\
0.91 & 0.74 & 3.73 & 3.70 & 0.01 & 0.0028 & 0.0023 & 0.32 \\
0.89 & 0.78 & 4.16 & 4.16 & 0.39 & 0.0023 & 0.0023 & 0.47 \\
0.89 & 0.82 & 3.76 & 3.76 & 0.12 & 0.0034 & 0.0029 & 0.15 \\
0.89 & 0.82 & 3.80 & 3.80 & 0.11 & 0.0034 & 0.0029 & 0.14 \\
0.87 & 0.78 & 4.12 & 4.08 & 0.03 & 0.0027 & 0.0027 & 0.01 \\
0.87 & 0.69 & 4.41 & 4.41 & 0.30 & 0.0026 & 0.0026 & 0.46 \\
0.84 & 0.74 & 4.54 & 4.17 & 0.49 & 0.0025 & 0.0021 & 0.49 \\
0.89 & 0.81 & 3.81 & 3.81 & 0.11 & 0.0057 & 0.0041 & 0.01 \\
0.88 & 0.80 & 3.87 & 3.87 & 0.11 & 0.0049 & 0.0038 & 0.05 \\
0.91 & 0.74 & 3.73 & 3.70 & 0.01 & 0.0092 & 0.0092 & 0.01 \\
0.80 & 0.74 & 4.51 & 4.49 & 0.40 & 0.0027 & 0.0027 & 0.25 \\
\rowcolor{LightCyan} 0.82 & 0.73 & 5.19 & 4.57 & 0.07 & 0.0023 & 0.0019 & 0.11 \\ 
0.76 & 0.67 & 4.34 & 4.30 & 0.08 & 0.0022 & 0.0021 & 0.08 \\
0.88 & 0.80 & 5.00 & 4.78 & 0.03 & 0.0030 & 0.0030 & 0.05 \\
0.88 & 0.80 & 3.92 & 3.92 & 0.10 & 0.0060 & 0.0043 & 0.01 \\
0.86 & 0.70 & 4.56 & 3.69 & 0.14 & 0.0034 & 0.0032 & 0.43 \\
0.81 & 0.73 & 5.13 & 5.14 & 0.01 & 0.0027 & 0.0028 & 0.20 \\ 
0.81 & 0.73 & 4.09 & 4.09 & 0.09 & 0.0045 & 0.0038 & 0.01 \\
0.80 & 0.74 & 4.51 & 4.49 & 0.40 & 0.0038 & 0.0038 & 0.42 \\
0.80 & 0.68 & 4.14 & 4.14 & 0.29 & 0.0040 & 0.0040 & 0.38 \\
0.80 & 0.66 & 4.81 & 4.81 & 0.21 & 0.0038 & 0.0038 & 0.10 \\
0.80 & 0.74 & 4.60 & 4.60 & 0.35 & 0.0041 & 0.0041 & 0.12 \\
0.88 & 0.69 & 4.98 & 4.98 & 0.34 & 0.0066 & 0.0066 & 0.27 \\
0.78 & 0.71 & 4.60 & 4.60 & 0.35 & 0.0041 & 0.0041 & 0.12 \\
0.79 & 0.69 & 6.68 & 5.54 & 0.05 & 0.0028 & 0.0023 & 0.32 \\
0.81 & 0.72 & 4.70 & 4.70 & 0.09 & 0.0070 & 0.0070 & 0.03 \\
0.77 & 0.65 & 4.83 & 4.83 & 0.20 & 0.0038 & 0.0038 & 0.07 \\
0.87 & 0.70 & 4.91 & 4.79 & 0.36 & 0.0073 & 0.0073 & 0.46 \\
0.77 & 0.68 & 5.85 & 5.85 & 0.38 & 0.0032 & 0.0032 & 0.34 \\
0.71 & 0.63 & 6.37 & 6.71 & 0.39 & 0.0026 & 0.0026 & 0.32 \\
0.71 & 0.63 & 6.37 & 6.74 & 0.25 & 0.0026 & 0.0026 & 0.27 \\
0.79 & 0.66 & 4.69 & 4.69 & 0.09 & 0.0093 & 0.0093 & 0.01 \\
0.76 & 0.69 & 5.00 & 4.98 & 0.35 & 0.0042 & 0.0042 & 0.38 \\
0.72 & 0.62 & 5.70 & 5.70 & 0.41 & 0.0036 & 0.0036 & 0.43 \\
0.73 & 0.59 & 9.63 & 9.63 & 0.45 & 0.0029 & 0.0029 & 0.48 \\
0.80 & 0.68 & 6.00 & 6.00 & 0.11 & 0.0075 & 0.0075 & 0.17 \\
0.71 & 0.58 & 9.70 & 9.70 & 0.40 & 0.0029 & 0.0029 & 0.26 \\
0.74 & 0.68 & 5.70 & 4.60 & 0.21 & 0.0070 & 0.0042 & 0.35 \\
0.74 & 0.68 & 5.70 & 4.60 & 0.21 & 0.0070 & 0.0042 & 0.35 \\
0.77 & 0.65 & 6.73 & 6.73 & 0.13 & 0.0094 & 0.0094 & 0.02 \\
0.78 & 0.68 & 7.39 & 7.39 & 0.12 & 0.0095 & 0.0095 & 0.04 \\
0.78 & 0.66 & 8.43 & 5.09 & 0.48 & 0.0096 & 0.0095 & 0.50 \\

\cmidrule(lr){1-8}

\end{tabularx}
\caption[]{Competition leaderboard table at 4 May 2018 where each row represents an entry from a competition team and our entry is highlighted.  There are three target outcomes for prediction: 1) Diagnosis, 2) the ADAS-13 score, and 3) VENTS-ICV which is the ventricles volume divided by intracranial volume. Predictions for diagnosis are presented as relative probabilities for each of the three potential diagnostic categories.  For the ADAS score, and the normalised ventricles volume, we provide confidence intervals which indicate in the limit where 50\% of the predictions would lie if an experiment were repeated many times on new data. The rank is determined by the lowest sum of ranks from mAUC, ADAS-13 MAE and VENTS-ICV MAE.   The metrics are based on those used in the TADPOLE competition.  For diagnosis classification we use the multiclass area under the receiver operating curve (mAUC) and balanced classification accuracy (BCA) as described in Table \ref{tab:results_diag}.  The metrics used for ADAS-13 and VENTS-ICV are the mean absolute error (MAE), the weighted error score (WES), which is the absolute error weighted by the inverse of the confidence interval range, and the coverage probability accuracy (CPA), defined as, $CPA = |j - 0.5|$, where $j$ is the proportion of measurements falling within the 50\% confidence interval.}
\label{tab:leaderboard}
\end{table}
%
\section{Discussion}
In selecting participants for clinical trials, a positive PET scan is commonly used as part of the inclusion criteria.  However PET imaging is expensive, so when a positive scan is one of the trial inclusion criteria it is desirable to avoid screening failures.  So one application of predicting Alzheimer's disease is to preselect candidates before applying the criteria.  Time series collected both in clinic and from studies such as ADNI inevitably have missing data points, and are of variable length.  The same will be true for clinical data collected from patients, since it is not uncommon for appointments to be missed, and for people to withdraw from data collection for various reasons.  Most time series prediction methods assume data which is complete and regularly sampled, so that it has to be pre-processed using imputation or interpolation methods to fulfil this assumption.  In this paper, rather than using a traditional time series method we have used a machine learning method to learn the relationship between pairs of time points at different separations.  The input vector comprises a summary of the time series history up to that point and the demographic and non-time varying factors such as genetic data. This method makes no assumptions about the dynamics of the time series, and it is applicable to data which has missing and irregularly sampled points.  The results are better than a baseline last-value estimator, and they validate the method as effective.

%
\section{Acknowledgements}
Data collection and sharing for this project was funded by the Alzheimer's Disease
Neuroimaging Initiative (ADNI) (National Institutes of Health Grant U01 AG024904) and
DOD ADNI (Department of Defense award number W81XWH-12-2-0012). A full list of funding
sources for ADNI is provided in the document `Alzheimer’s Disease Neuroimaging Initiative (ADNI)
Data Sharing and Publication Policy' available through \url{adni.loni.usc.edu/}.  
\nind
This work uses the TADPOLE data sets https://tadpole.grand-challenge.org constructed by the EuroPOND consortium http://europond.eu funded by the European Union’s Horizon 2020 research and innovation programme under grant agreement No 666992.  
\nind
The MRC Dementias Platform UK (DPUK) \url{https://www.dementiasplatform.uk/} provided support in the preparation of this paper.  DPUK is a multi-million pound public-private partnership, developed and led by the MRC, to accelerate progress in and open up dementias research. The aims of DPUK are early detection, improved treatment and ultimately the prevention of dementias.
%
\section*{}

\bibliography{main}

\end{document}

%% file: macpap2.tex
\newcommand{\be}{\begin{equation}}
\newcommand{\ee}{\end{equation}}
\newcommand{\bean}{\begin{eqnarray}}
\newcommand{\eean}{\end{eqnarray}}
\newcommand{\bea}{\begin{eqnarray*}}
\newcommand{\eea}{\end{eqnarray*}}
\newcommand{\bfig}{\begin{figure}}
\newcommand{\efig}{\end{figure}}
\newcommand{\ba}{\begin{array}}
\newcommand{\ea}{\end{array}}

\newcommand{\eps}{\epsilon}

\newcommand{\nind}{\newline\indent}

\newcommand{\etal}{\emph{et al. }}
\newcommand{\etalns}{\emph{et al.}}
